\documentclass[runningheads]{llncs}

\usepackage[T1]{fontenc}
\usepackage{graphicx}
\usepackage{amssymb}
\usepackage{amsmath}

\usepackage[hyphens]{url}
\usepackage{hyperref}
\usepackage[hyphenbreaks]{breakurl}

\usepackage{cite}

\usepackage{microtype}
\usepackage{mathtools}

\DeclarePairedDelimiter\floor{\lfloor}{\rfloor}

\begin{document}

\title{Maximizing the practical achievability of quantum annealing attacks on factorization-based cryptography}
%
%
\author{Olgierd Żołnierczyk\inst{1}\orcidID{0000-0002-5196-3494}}
\authorrunning{Olgierd Żołnierczyk}
%
\institute{Military University of Technology, Warsaw, Poland, 
\email{olgierd.zolnierczyk@wat.edu.pl}\\ }
\maketitle              
\begin{abstract}
This work focuses on quantum methods for cryptanalysis of schemes based on the integer factorization problem and the discrete logarithm problem. We demonstrate how to practically solve the largest instances of the factorization problem by improving an approach that combines quantum and classical computations, assuming the use of the best publicly available special-class quantum computer: the quantum annealer. We achieve new computational experiment results by solving the largest instance of the factorization problem ever announced as solved using quantum annealing, with a size of 29 bits. The core idea of the improved approach is to leverage known sub-exponential classical method  to break the problem down into many smaller computations and perform the most critical ones on a quantum computer. This approach does not reduce the complexity class, but it assesses the pragmatic capabilities of an attacker. It also marks a step forward in the development of hybrid methods, which in practice may surpass classical methods in terms of efficiency sooner than purely quantum computations will.

\end{abstract}
%
%
%

\section{Introduction}

Cryptoschemes, which are at most as secure as the factorization problem  (it means factorization-based), can be compromised in standard, general approaches: classically with subexponential complexity through decomposition base methods \cite{Crandall2005} and quantumly with polynomial complexity through Shor's algorithm \cite{Shor94} or via other quantum methods of unspecified complexity. Importantly, fully quantum solutions in practice require too many resources: the largest practically solved examples include: 6 bits - Shor's algorithm \cite{Amico2019}, 10 bits - adiabatic quantum computation \cite{Pal2019}, and 23 bits - quantum annealing \cite{Ding2024}. 
Therefore, hybrid methods such as \cite{Bernstein2017, Pal2019, Wronski2021, Zolnierczyk2023} have been invented, representing a compromise between computational complexity and the use of quantum resources.

In this work, we present experimental results of the efficiency of hybrid classical-quantum computations using quantum annealing. This is a transitional approach  between fully quantum methods and best methods for classical hardware. It involves factoring numbers based on a subexponential factorization method, where part of the computations is performed using quantum techniques, specifically through quantum annealing. This phase is crucial for the complexity of the method and focuses on searching for numbers with special properties, known as $B$-smooth numbers. As a result, we obtain an improvement that has the potential to practically address the largest instance, ever computed on quantum machine. This is due to using fewer quantum resources than fully quantum approaches.

 As a result, a 29-bit problem was solved, setting a record among all approaches with quantum annealing. This result improves upon the 23-bit result of a fully quantum annealing method, and the most effective to date, a hybrid approach, which factored a 26-bit number \cite{Zolnierczyk2023}. The contribution of this work to improving this record lies in using a more advanced algebraically decomposition base method (General Number Field Sieve) for factorization compared to \cite{Zolnierczyk2023}. The quantum resource occupancy $O(\frac{\log^{2}{h}}{4})$ remains the same for $h$ - values from the relation gathering step, yet $h$ are smaller in this work due to the refined algebraic sieving of the number field. Among all hybrid approaches, this also represents the largest result, excluding only \cite{Karamlou2021}, where the type of scalability is unspecified. 

Due to the lack of theoretical foundations (regarding the complexity of quantum annealing), the aim of this paper is not to fully compare this approach with known ones, including classical subexponential solutions.
 Although the complexity of presented calculations remains in the subexponential class, in practice, hybrid approaches may surpass classical solutions sooner than fully quantum methods

\subsubsection{Paper organization}
The article is divided into two main parts: the established theory and the description of the results of this work – the experiment. The following sections explain:
\begin{itemize}
\item \autoref{GNFS} The method of the General Number Field Sieve and the role of finding $B$-smooth numbers within.
\item \autoref{sec:QAintro} What is quantum annealing.
\item \autoref{sec:QA}  How quantum annealing can be used to find $B$-smooth numbers.
\item \autoref{sec:experiment} The contribution of this work, how the experiment was conducted, and the detailed results obtained.
\end{itemize}

\section{Theory}\label{sec.gnfs-annealing}

 \subsection{Classical factorization method}\label{GNFS}

In cryptology integer factorization goal is to factor  $N = pq$, where $p,q$ are primes. The most effective classical general method for integer factorization  is the General Number Field Sieve (GNFS), initially proposed as a generalized version of previous ideas in \cite{Lenstra1990}.  There are many improvements  of this method; however, from the perspective of this work, it is sufficient to focus on the primary version.  It  achieves subexponential cost 
\begin{equation*}
    O\left(\exp{\left(\left(1.923 + o(1)\right)(\ln{N})^{1/3}(\ln{\ln{N}})^{2/3}\right)}\right),
\end{equation*}
under various heuristic assumptions.

\subsubsection{Essence of GNFS} This tool is highly sophisticated and operationalises the fundamental idea of factorization: establishing relationships of congruences of squares, defined as follows

\begin{equation}\label{eq:GNFSMainRelation}
\begin{aligned}
 X^{2} &\equiv Y^{2} \pmod{N}.  
\end{aligned}
\end{equation}
If we can indicate such a pair, we then have: $(X-Y)(X+Y) \equiv 0 \pmod{N}$, and if $X \not \equiv \pm Y \pmod{N}$, we can easily factorize $N$ by computing $\gcd(X \pm Y, N)$.  Establishing relation from \autoref{eq:GNFSMainRelation} leads to factorization with  probability at least $\frac{1}{2}$. This is accomplished through the following stages. 

Firstly, finding an appropriate polynomial $\mathcal{F}(x) \in \mathbb{Z}[x]$, and an $m$ such that
\begin{equation}\label{eq:polf}
    \mathcal{F}(m) \equiv 0 \pmod{N}.
\end{equation}
The simplest way to do this is by: setting \( m = \lfloor N^{\frac{1}{d}} \rfloor \), where \( d \) is an integer parameter that regulates the degree of the polynomial \( \mathcal{F} \), then expressing \( N \) in base \( m \), that is:
$N = m^{d} + y_{d-1}m^{d-1} + \dots + y_{1}m + y_{0}$,
where each of \( y_{d-1}, \dots, y_{1}, y_{0} \) is between 0 and \( m-1 \) (for small $d$ we have $(N > 2m^{d})$), and finally assigning \( \mathcal{F} = x^{d} + y_{d-1}x^{d-1} + \dots + y_{1}x + y_{0} \). Hence, it follows \autoref{eq:polf}. We assume that $F$ is irreducible, because otherwise we immediately obtain the factorization of $N$. Moreover, if we denote $\rho$ as a root of $\mathcal{F}$, this polynomial defines a ring homomorphism (proof in \cite{Jarvis2014}, Lemma. 11.19):
\begin{equation}\label{eq:phiDefinition}
\begin{aligned}
&\phi: \mathbb{Z}[\rho]  \rightarrow \mathbb{Z}/N\mathbb{Z},\\
&\sum_{i} z_{i} \rho^{i} \rightarrow  \sum_{i}z_{i}m^{i}  \pmod{N},
\end{aligned}
\end{equation}
and let us recall that a homomorphism, by definition, has the following defining property:
\begin{equation}\label{eq:homProperty}
    \forall_{\gamma_{1} \gamma_{2} \in \mathbb{Z}[\rho]} \phi \left( \gamma_{1} \gamma_{2}\right) = \phi \left( \gamma_{1} \right) \phi \left( \gamma_{2} \right) \in  \mathbb{Z}/N\mathbb{Z}.
\end{equation}

Secondly, identifying a set $\mathbb{S}$ of pairs $(a,b)$ such that combined: ${\displaystyle  \prod_{(a,b) \in \mathbb{S}} (a+b\rho)}$ is a square $\gamma^{2} \in \mathbb{Z}[\rho]$ and ${\displaystyle \prod_{(a,b) \in S}} (a+bm)$ is a square $X^{2} \in \mathbb{Z}$. 

Third, by computing  $ \gamma \in  \mathbb{Z}[\rho] $ as square root of $\gamma^{2}$. At this point, the factorization is almost complete because the following holds: 
\begin{equation}
\begin{split}
X^{2} = \prod_{(a,b) \in \mathbb{S} }(a+bm) &\overset{\text{ by \autoref{eq:phiDefinition}}}{\underset{\pmod{N}}{\equiv}} \phi \left(Y^{2}\right)  \\  &\overset{\text{by \autoref{eq:homProperty}}}{=} \phi(\gamma)\phi(\gamma) \; \text{mod N}\\ &\;\;\;\;\;\;= Y^{2} \; \text{mod N},
\end{split}
\end{equation}
for some $Y \in  \mathbb{Z}/N\mathbb{Z}$ (the symbol $\underset{\pmod{N}}{\equiv}$ here denotes congruence modulo $N$).  
Hence, we have relation from \autoref{eq:GNFSMainRelation} and can easily obtain $X$ by taking the square root of $X^{2}$ in $\mathbb{Z}$ (and then reducing modulo $N$), and  $Y$ by projecting $\phi(\gamma)$. 
Equally important, when computing square roots in these two rings, there is no map that guarantees $X \equiv \pm Y \pmod{N}$, and in practice, this is often not the case. Then, we satisfy the conditions to determine a non-trivial divisor of $N$ by computing $\gcd(X-Y, N)$ or $\gcd(X+Y, N)$. This is the outline of the entire method.

All three phases contain more solutions that address the challenges associated with each of them. However, in particular, the most important stage for this work and for the complexity of the entire method is the stage of determining the relation of congruent squares.

\subsubsection{Linear algebra stage} The forms $\gamma^{2}$ and $X^{2}$ (being products of many simple expressions) are not insignificant here. This arises, among other things, from the fact that instead of checking each randomly selected pair belonging to $\mathbb{Z} \times \mathbb{Z}[\rho]$ to see whether it forms a pair of squares (a naive approach), we determine the desired square values by multiplying the appropriate combination of previously recorded simple expressions, represented by $(a,b)$. This multiplication is understood as the multiplication of the expressions of the form $a+bm$ and $a + \rho m$ in their respective two rings. Thus, we treat this specific set of pairs $(a,b)$ as a single representation of both elements from the two rings, as we need to obtain squares in both rings simultaneously. 

If we succeed in identifying a sufficient number of pairs $(a,b)$ that imply only those elements that can be expressed as a unique factorization into prime elements (this applies to both integers and the ring $\mathbb{Z}[\rho]$, in which we factor the element $\gamma^{2}$), we reduce the problem of obtaining congruent squares to solving a system of linear equations over the field $\mathbb{F}_{2}$. The assumption of unique factorization is unrealistic in $\mathbb{Z}[\rho]$, but it is close enough to practice to clearly present the main idea of this phase; hence, we leave the realistic refinements for later.

To illustrate this more clearly, let us assume we represent a fixed pair $(a_{1}, b_{1})$. We record the factorization in $\mathbb{Z}$: $a_{1}+b_{1}m = s_{1}^{e_{1}} s_{2}^{e_{2}} s_{3}^{e_{3}} \dots s_{k}^{e_{k}}$, and the factorization in the ring of algebraic integers of the number field $\mathbb{Q}(\rho)$: $a_{1}+b_{1}\rho  = \mathfrak{p}_{1}^{v_{1}}, \mathfrak{p}_{2}^{v_{2}}, \mathfrak{p}_{3}^{v_{3}} \dots \mathfrak{p}_{t}^{v_{t}}$. Now, the sequence $e_{1}e_{2}e_{3} \dots e_{k}v_{1}v_{2}v_{3}v_{t}$ is a vector representing the pair $(a_{1}, b_{1})$, which we shall call $\vec{V}_{1}$. Furthermore, the assumption of unique factorization still applies. We know that, in such a case, the parity of all exponents in any vector $\vec{V}$ is equivalent to the fact that the numbers it represents are squares. Having many different vectors $\vec{V}$, representing the recorded pairs $(a,b)$, we must determine their combination such that the sum of the coefficients of these vectors for each dimension is even. This implies linear vector calculations over $\mathbb{F}_{2}$:
\begin{equation}
\mathrm{x}_{1}V_{1} + \mathrm{x}_{2}V_{2} + \dots \mathrm{x}_{l}V_{l} =  \vec{0}.
\end{equation}
The vector $\mathrm{x}_{1}, \mathrm{x}_{2}, \dots , \mathrm{x}_{l}$, where $\mathrm{x}_{i} \in \mathbb{F}_{2}$, encodes the solution to the system of linear equations defined by the matrix formed from the vectors $\vec{V}$, thereby indicating which pairs $(a,b)$ should be multiplied together so that the result is a square in both rings.

\subsubsection{Factorization of algebraic elements} In practice, to establish the vector representation of the screened pairs $(a,b)$, we must determine a finite set of elements $s$ and $\mathfrak{p}$ necessary for this, which we call the factor base $\mathbb{B}$. For integers, we rely on the unique factorization of integers and use the set of prime numbers smaller than a given bound $B$, which gives rise to the term $B$-smooth number, meaning 
\begin{definition}
    A $B$-smooth number is one whose prime divisors are all less than or equal to $B$.
\end{definition}
In the second structure, we do not have unique factorization; instead, we use the unique factorization into so-called prime ideals in the ring of integers of the number field $\mathbb{Q}(\rho)$ (in this case, the smoothness of the screened algebraic elements also depends on the bound $B$, as described more thoroughly below).

All the concepts and facts from number field theory necessary for a complete understanding of the factorization of algebraic elements (from the ring $\mathbb{Z}[\rho]$) can be found in \cite{Jarvis2014}. From the perspective of this work, two issues are important: firstly, how factorization into prime ideals relates to detecting $B$-smoothness, and secondly, what problems it poses and what practical steps are involved, ultimately aimed in this phase at determining the pair $\gamma^{2}, X^{2}$. The use of factorization into prime ideals makes verifying the smoothness of an algebraic element (i.e., confirming that the element factors into prime ideals from the factor base) more complex and requires a two-step process.

This is because:
\begin{theorem}
The factorization into prime ideals of the ideal generated by $(a + b\rho)$, assuming that $\gcd(a,b) = 1$, contains only ideals generated by the pair: $(s, \rho -r)$, where:
\begin{itemize}
\item $s$ is a prime number such that $\mathcal{F}$ has roots modulo $s$,
\item $r$ satisfies: $\mathcal{F}(r) \equiv 0 \pmod{s}$.
\end{itemize}
\end{theorem}
(The proof of this theorem can be derived based on the proof in \cite{Jarvis2014}, Theorem 11.12). Therefore, the prime ideals from the factor base can be represented by pairs $(L,r)$. Hence, the factor base is narrowed down to those prime ideals generated by primes from the factor base $\mathbb{B}$ that can occur in the factorization of $a + b\rho$, according to the theorem. Thus, the first and most important step in verifying the smoothness of an algebraic element is identifying all prime numbers that generate the prime ideals of that element. This is made possible by using the norm of the number field element, which intuitively has the following meaning: it generalizes the absolute value of a real number and serves as a measure of a certain type of magnitude of a given element. It is defined as follows:

\begin{definition}{}
    The norm of an element $\alpha$ of the number field $\mathbb{Q}(\rho)$, denoted $N_{\mathbb{Q}(\rho)}(\alpha)$, is:
\begin{equation}
N_{\mathbb{Q}(\rho)}(\alpha) = \prod_{i}\sigma_{i}(\alpha),
\end{equation}
where $\sigma_{i}$ is the i-th field homomorphism, defined as:

\begin{equation}\label{eq:embedding}
\begin{aligned}
&\sigma_{i}: \mathbb{Q}(\rho)  \rightarrow \mathbb{Q}(\rho_{i}),\\
&\sum_{j} \mu_{j} \rho^{j} \rightarrow  \sum_{j}\mu_{j}\rho_{i}^{j}  \pmod{N},
\end{aligned}
\end{equation}

and $\rho_{i}$ is the i-th root of the minimal polynomial of $\rho$ in the field $\mathbb{C}$.
\end{definition}

Additionally, it turns out that:

\begin{theorem}
If $\alpha$ belongs to the ring of integers of the number field $\mathbb{Q}(\rho)$, then $N_{\mathbb{Q}(\rho)}(\alpha) \in \mathbb{Z}$.
\end{theorem}

(The proof of this theorem can be found in \cite{Jarvis2014}, Corollary 3.17), therefore in this case, we can discuss the prime factorization of the norm, especially since we can compute it efficiently, as:

\begin{align*}
N_{\mathbb{Q}(\rho)}(a + b\rho) &= (a + b\rho)(a + b\rho_2) \cdots (a + b\rho_d) \\
&= (-b)^d \left( \frac{a}{b} - \rho \right) \left( \frac{a}{b} - \rho_2 \right) \cdots \left( \frac{a}{b} - \rho_d \right) \\
&= (-b)^d \mathcal{F}\left( \frac{a}{b} \right).
\end{align*}

Adding to this the even more significant fact that:

\begin{theorem}
A prime number $s$ divides $N_{\mathbb{Q}(\rho)}(\alpha)$ if and only if the ideal generated by $\alpha$ includes in its factorization a prime ideal generated by the pair $(s, r)$.
\end{theorem}
(The proof of this theorem can be found in \cite{Buhler2006}, Preposition 5.3.), we can conclude that the determination of all prime ideals into which an algebraic element factors is performed in the first step by factoring the norm of elements of the form $a + b\rho$, where $\gcd(a,b) = 1$, into prime numbers. To fully represent these ideals, a second step is required — the determination of the corresponding $r$, which is done via simple linear modular calculations. For clarity, we will omit these calculations and refer to a more comprehensive description, along with other details of the GNFS method, in \cite{lenstra1993development}.

Even after determining the exponents, i.e., the vectors $\vec{V}$, the situation is not as straightforward as under the idealised assumption of unique factorization in $\mathbb{Z}[\rho]$. Factorization into ideals introduces two additional potential points of failure related to representing the element by an ideal and vice versa, as well as two more points of failure associated with the fact that the ideals are defined not in $\mathbb{Z}[\rho]$ but in the ring of integers of the number field $\mathbb{Q}(\rho)$. In practice, we deal with these issues by determining additional properties for each pair $(a,b)$, in the form of extra columns in the matrix $M$ (referred to as Adleman columns). These do not guarantee the success of identifying the correct pair of squares, but they significantly increase the probability of success to a sufficiently high level.

\subsubsection{Closing notes} The remaining phases of the GNFS method, namely the selection of the polynomial $\mathcal{F}$ and the determination of the square root of $\gamma^{2}$, can be approached using various methods, which have been developed and refined over time by researchers. Further details on these phases, as well as improvements in the sieving process, can be found in works such as \cite{Guillevic2021, thome2012, Bouillaguet2023}.

Crucially for this work, the factorization problem is effectively broken down into subproblems involving determining whether a given element is divisible by prime factors greater than a certain limit $B$ (there is also the cost of additional computations, which does not increase the complexity of the entire method.). We say that we are checking the $B$-smoothness of a given number. Thus, identifying a sufficient amount of $B$-smooth numbers among those sieved provides a high probability of factorizing the $N$.

\subsection{Quantum annealing and the QUBO problem}\label{sec:QAintro}
The most general concept of quantum annealing, introduced in \cite{Kadowaki1998}, is that due to the quantum phenomena involved, the computations performed by a quantum annealer rely on solving a certain type of optimization problems, defined below. In this process, the minimum of a given objective function $\mathcal{K}$ is found with some probability, where the preimage of the function is represented by the state of the quantum computer's memory (the quantum annealer). From a physical perspective, the computations involve reaching the system's lowest energy level (ground state), which is achievable through quantum phenomena such as superposition, tunneling, and energy dissipation. This is a computational technology derived from the paradigm of adiabatic computing \cite{vanDam2001,farhi2000}. 


In practice, a quantum annealing computer resolves computational tasks defined by the Ising problem, which can equivalently be transformed (through a linear transformation) into the form of Quadratic Unconstrained Binary Optimization (QUBO). QUBO is a second-degree, multivariate polynomial with real coefficients and binary variables:
\begin{equation}\label{eq:qubo}
\mathcal{Q}(\mathrm{u_{1}},\dots,\mathrm{u_{o}}) = \sum_{i}\beta_{i}\mathrm{u_{i}} + \sum_{i<j}\delta_{i,j}\mathrm{u_{i}}\mathrm{u_{j}}.
\end{equation}

 Transforming any problem into the QUBO  allows for attempts to solve it via quantum annealing.  

 It has already been shown how to transform into the QUBO form: the discrete logarithm problem over a finite field \cite{Wroński2022}, as well as the discrete logarithm problem in the group of points on an elliptic curve \cite{Wroński2024}, block cipher equations \cite{Burek2023, Burek2022}, and stream cipher equations \cite{Wroński2023/2, Leniak2024}.

 Some results suggest that the time complexity of quantum annealing falls within the sub-exponential class; however, so far, determining the full formal complexity requires further research. Another important issue related to the specification of quantum annealing is that the quantum annealing qubit  should not be compared to the qubit  in gate-based quantum computers, as these are different types of hardware components.

\subsection{Quantum annealing methods for factorization and determining B-smooth Numbers}\label{sec:QA}

\subsubsection{Direct factorization}
Direct methods for factorization of $N = pq$ through quantum annealing are known for several years, see \cite{Jia18, Pen19, Mengoni2020}.  The main idea of transforming the problem involves the binary representation $\mathrm{p_{\tau(p)}}\dots \mathrm{p_{3}}\mathrm{p_{2}}\mathrm{p_{1}}\mathrm{p_{0}}$ of $p$, the binary representation $\mathrm{q_{\tau(q)}}\dots \mathrm{q_{3}}\mathrm{q_{2}}\mathrm{q_{1}}\mathrm{q_{0}}$ of $q$ and defining a cost function of these variables

\begin{equation}
    \begin{split}
        &\underset{_{dir}}{\mathcal{K}}\left(\mathrm{p_{0}},\mathrm{p_{1}}, \mathrm{p_{2}}, \dots, \mathrm{p_{\tau(p)}}, \mathrm{q_{0}},\mathrm{q_{1}}, \mathrm{q_{2}}, \dots \mathrm{q_{\tau(q)}}\right) = \\
& \left(N - \left(\sum_{i} \mathrm{p_{i}}2^{i}\right)\left(\sum_{i} \mathrm{q_{i}}2^{i}\right)\right)^{2} = \left(N - pq\right)^{2},
    \end{split}
\end{equation}

which gives us the optimization form of the factorization problem, but not the QUBO form.

\subsubsection{Degree reduction of a polynomial} To reduce $\underset{_{dir}}{\mathcal{K}}$ to the QUBO form $\underset{_{dir}}{\mathcal{Q}}$, we must reduce the degree of the polynomial $\underset{_{dir}}{\mathcal{K}}$ to 2. The reduction process involves applying a series of transformations $\lambda$ to each monomial of the original function to obtain a new function, which is a polynomial of the appropriate degree. The transformation $\lambda$ is defined to preserve the minima of the function at the same points (values of the arguments) but reduces the degree of the given monomial. The transformation $\lambda$ is performed by introducing a new auxiliary variable $\hat{\mathrm{a}}\in \{0,1\}$ and adding a so-called penalty $\mathcal{P}_{\hat{\mathrm{a}}}$:
\begin{equation}
\begin{split}
\mathrm{p_{0}}\mathrm{p_{1}}\mathrm{p_{2}} \rightarrow \hat{\mathrm{a}}\mathrm{p_{1}} + \mathcal{P}_{\hat{\mathrm{a}}},\\
\mathcal{P}_{\hat{\mathrm{a}}} = 2(\mathrm{p_{0}}\mathrm{p_{1}} - 2\hat{\mathrm{a}}(\mathrm{p_{0}} + \mathrm{p_{1}}) + 3\hat{\mathrm{a}})
\end{split}
\end{equation}
Considering the above expressions as functions, we find that the minimal values of the expressions (i.e., zero) on both sides of the transformation are achieved for the same values of $\mathrm{p}_{0},\mathrm{p}_{1},\mathrm{p}_{2}$, which can be easily verified through exhaustive search. This also holds true for expressions of the form $\psi \mathrm{p}_{0}\mathrm{p}_{1}\mathrm{p}_{2}$, where $\psi \in \mathbb{R}$ (thus for any monomial form), and obviously remains true for functions that are sums of such monomials.

The reduction can be performed on the form of the function before squaring (in which case we must reduce the degree of the polynomial to 1, linearization see \cite{Zolnierczyk2023} ), and we denote it by $\mathring{\lambda}_{lin}$ or after squaring (in which case it is sufficient to reduce the degree to 2, quadratization see \cite{Jia18}), and we denote it by $\mathring{\lambda}_{quad}$.

\subsubsection{Detecting smoothness}\label{sec:detecting}
Factorization through quantum annealing is used to detect $B$-smooth numbers, whose identification is motivated as described in  \autoref{GNFS}. This is done in a very intuitive and simple way. A potentially smooth number $h > B$ is broken down into two divisors 
\begin{equation}
h = f g.
\end{equation}
We define the lengths: $\tau(g)$ as the number of bits $\floor{h/s_{1}}$ and $\tau(f)$ as the number of bits $s_{k}$, where $s_{1}$ is the smallest and $s_{k}$ is the largest prime in the chosen base $\mathbb{B}$. Additionally, instead of taking $s_{1}$ as a reference, we can choose a larger number from the base and check and record the divisibility of $h$ by the smallest primes (for example $2, 3$) in an efficient manner before the annealing process, which can increase the efficiency of the procedure.

The values $\tau(g)$ and $\tau(f)$ determine the number of binary variables used to represent $g$ and $f$, respectively. If we obtain a non-trivial $f \mid h$, we can proceed to further factorize $g$ (if necessary), recursively invoking the procedure until all divisors are less than or equal to $B$ and we confirm smoothness. Otherwise, we treat the number $h$ as non-smooth. This results in a simple procedure for finding $B$-smooth numbers through quantum annealing.

\subsubsection{Multiplication table procedure}
In practice, for smoothness detection, we use an enhanced method of factorization through quantum annealing. Specifically, a multiplication table is established, constructed exactly according to the long multiplication technique, where the binary variables of the table are $h_{i}, f_{i}, g_{i}$, representing the bits of $h, f, g$, respectively, and the carry bits $c_{i}$. The table is then divided into blocks, each covering several columns. The width of the $i$-th block (i.e., the number of its columns) is denoted by $w_{i}$. This gives us a  \autoref{multiplication}, where $w_{0}$ is always equal to 1, because, in practice, checking for divisibility by at least $s_{1} = 2$ occurs before the smoothness verification procedure. 

\begin{table}[ht]
\centering
 \resizebox{\columnwidth}{!}{\begin{tabular}{||c c c c c c c c c c | c c | c ||} 
 \hline
  &&  & & & &$\mathrm{f_{\tau(f)}}$  &$.$ &$.$ &$.$ & $\mathrm{f_{2}}$&$\mathrm{f_{1}}$& $1$ \\ [0.5ex] 
  &&  & & & &$\mathrm{g_{\tau(g)}}$ &$.$ &$.$  &$.$& $\mathrm{g_{2}}$&$\mathrm{g_{1}}$& $1$ \\ [0.5ex] 
 \hline\hline
  &&  && &$c_{0}$ &$\mathrm{f_{\tau(f)}}$ &$.$ &$.$ & $.$& $\mathrm{f_{2}}$    & $\mathrm{f_{1}}$   & $1$ \\ 
  &&  &&$c_{1}$&$\mathrm{f_{\tau(f)}}\mathrm{g_{1}}$ &$.$ &$.$ &$.$ & $\mathrm{f_{2}}\mathrm{g_{1}}$ & $\mathrm{f_{1}}\mathrm{g_{1}}$& $\mathrm{g_{1}}$&\\
  &&  &$.$&$\mathrm{f_{\tau(f)}}\mathrm{g_{2}}$&$.$ &$.$  &$.$ &$\mathrm{f_{2}}\mathrm{g_{2}}$&$\mathrm{f_{1}}\mathrm{g_{2}}$&$\mathrm{g_{2}}$&&  \\
  &&$.$  &$.$&& & & & & $.$   &     &    &    \\
  &$.$&$.$  &&& & & &$.$  &   &     &    &    \\
  $c_{k}$&$.$&&  & & & & $.$&  &   &     &    &     \\
 $\mathrm{f_{\tau(f)}}\mathrm{g_{\tau(g)}}$&$.$&$.$ &$.$&$\mathrm{f_{2}}\mathrm{g_{\tau(g)}}$ &$\mathrm{f_{1}}\mathrm{g_{\tau(g)}}$ &$\mathrm{g_{\tau(g)}}$&  &  & & & &\\[1ex]
 \hline \hline
  $\mathrm{h_{\tau(h)}}$& & & &   $.$ & &$.$ & &$.$ & & $\mathrm{h_{2}}$ & $\mathrm{h_{1}}$ & $1$ \\ [0.5ex]
  \hline
  \multicolumn{2}{c}{\upbracefill}&\multicolumn{2}{c}{} &\multicolumn{1}{c}{.}&\multicolumn{1}{c}{}&\multicolumn{1}{c}{.}&\multicolumn{1}{c}{}&\multicolumn{1}{c}{.}&\multicolumn{1}{c}{}& \multicolumn{2}{c}{\upbracefill}&\multicolumn{1}{c}{\upbracefill}\\[-1ex]
  \multicolumn{1}{c}{\scriptsize L-th block}&\multicolumn{9}{c}{} &\multicolumn{2}{c}{\scriptsize 1. block} &\multicolumn{1}{c}{\scriptsize 0. block}\\
 \end{tabular}}
 \caption{\label{multiplication} Multiplication table.}
 \end{table}

Instead of forming a single objective function $\mathcal{K}$ as before, we break down the problem by equating each expression $\mathcal{K}_{i}$ from the $i$-th block separately to the value read from the fragment of bits of the number $h$, as an independent binary representation: $n_{i}$. More specifically, we define $K_{i}$ as follows:

\begin{equation}
\begin{split}
  \mathcal{K}_{i} = \; &\overline{\mathcal{K}}_{i}(\mathrm{f_{0}},\mathrm{f_{1}}, \dots, \mathrm{f_{\tau(f)}}, \mathrm{g_{0}}, \mathrm{g_{1}}, \dots, \mathrm{g_{\tau(g)}} ) \\ &+ \mathcal{C}_{i}(\mathrm{c_{0}},\mathrm{c_{1}}, \dots, \mathrm{c_{J}}) - 2^{w_{i}}\mathcal{C}_{i+1}(\mathrm{c_{0}},\mathrm{c_{1}}, \dots, \mathrm{c_{J}}) - n_{i},  
\end{split}
\end{equation}
where:
\begin{itemize}
    \item[-] $\overline{\mathcal{K}}_{i}(\mathrm{f_{0}},\mathrm{f_{1}}, \dots, \mathrm{f_{\tau(f)}}, \mathrm{g_{0}}, \mathrm{g_{1}}, \dots, \mathrm{g_{\tau(g)}} )$ is the result of multiplying the bits $\mathrm{f_{i}}$ and $\mathrm{g_{i}}$ from the $i$-th block,
    \item[-] $C_{i}(\mathrm{c_{0}},\mathrm{c_{1}}, \dots, \mathrm{c_{J}})$ is an expression created by the shifted bits $c_{i}$ in the $i$-th block,
    \item[-] $\tau(c)$ is the number of carry bits into the $i$-th block.
\end{itemize}

In this way, we can formulate the QUBO form:
\begin{equation}
\mathcal{Q} = \mathring{\lambda}_{lin}\left(\mathcal{K}_{1}\right)^{2} + \mathring{\lambda}_{lin}\left(\mathcal{K}_{2}\right)^{2} +  \dots + \mathring{\lambda}_{lin}\left(\mathcal{K}_{L}\right)^{2}
\end{equation}

Additional improvements involve techniques for determining the total number of carry bits $J+1$, which have been explained more precisely in \cite{Zolnierczyk2023}.

Finally, quantum annealing is used to find the solution to the problem $Q$. By applying this procedure to the steps described in \autoref{sec:detecting}, we determine whether the number $h$ is $B$-smooth.


\section{Experiment}\label{sec:experiment}

\subsection{Contribution}
The contribution of this work lies in proposing an improvement to the hybrid quadratic sieve method, known from \cite{Zolnierczyk2023}, allowing for the assessment of the maximum potential of quantum annealing-based attacks on the integer factorization problem, and obtaining practical results through computational experiments.

The improvement involves the use of the known method for finding $B$-smooth numbers via quantum annealing, generally described in \autoref{sec:QA}, to sieve pairs $(a,b)$ in search of smooth elements in the GNFS method, as described in \autoref{sec.gnfs-annealing} (in the previous work, a less effective method – the quadratic sieve – was used). While the sieving was conducted through quantum computations, all other steps, such as polynomial selection, solving the system of linear equations, square root extraction, and smaller calculations, were performed classically (without a quantum computer). The sieving of pairs $(a,b)$ was done by attempting a quantum factorization of the corresponding values (see \autoref{GNFS}), preceded by a classical removal of the highest power of 2 dividing the value.

This improvement, which enabled achieving the maximum problem size, meant that the framework of hybrid (quantum-classical) computations, the most efficient of the known classical methods was utilized, while solutions to subproblems (searching for $B$-smooth relations) were sought through quantum annealing. This did not change the number of basic operations, thus the approach remains within the subexponential complexity class. The potential practical speed-up in full-scale attacks remains an unresolved issue due to the lack of precise knowledge about the complexity of quantum annealing. The answer to the question of whether the presented method will solve larger instances than the classical GNFS in full-scale applications lies in comparing the smoothness verification procedures in the classical and hybrid methods. Unfortunately, such a comparison is not possible when relying solely on computation time without determining the complexity of quantum annealing.

This practical example demonstrates how solving small subproblems by currently available quantum annealing computers, using all available results, realistically translates to the size of the factorization problem being solved. These are the implications of the possibility of a quantum solution for subproblems of this size.

The practical results of the experiment are described below.

\subsection{Purpose of the tests}
The goal of the experiment was to investigate the maximum size of the integer factorization problem in cryptography that can be solved using quantum annealing. This approach did not exclude the use of the best tools, both quantum and classical. By framing the problem in this way, we demonstrate how effectively cryptography based on the integer factorization problem and the discrete logarithm can already be attacked using quantum computers, despite remaining in the same computational complexity class, which, for now, cannot be practically changed. This provides a very realistic and current view of the range of possibilities available to attackers, while also taking a practical step toward hybrid attack ideas with lower complexity than classical methods, such as those proposed in \cite{Bernstein2017}.

\subsection{Methodology - classical part}
The methodology of the experiment comprises: detailed solutions in GNFS, the method of selecting a computational example, and the allowable precision for annealing. 
Key details of the GNFS method pertain to the implementation of random relation sampling, examined for B-smoothness. In the experiment, the pairs $(a,b)$ were limited by a constant constraint $0 < b \le D$, and a restriction $|a| \le A$, where $A$ was increased when, despite exhausting available pairs, no solution to the system of equations existed, according to \cite{Crandall2005}. The degree $d$ of the polynomial $\mathcal{F}$ was specified at the input, and a standard method for determining the polynomial $\mathcal{F}$ was adopted by expressing $N$ in base $m$. The search for examples was designed to limit the sizes of the algebraic-side value: $N_{\mathbb{Q}[\rho]}(a+bm)$ and the integer-side value $a+bm$, to be written on $W$ bits.  Importantly, the value of $W$ reflects the asymptotic size of the sieved numbers  to $N^{\frac{1}{d}}$, where $d \sim \sqrt[3]{\frac{3\log{N}}{\log{\log{N}}}}$ (see \cite{Jarvis2014}).
Examples exceeding this limit in the relation search phase were rejected. This naturally results from the computational limitations assumed for quantum annealing.

\subsection{Methodology - quantum part}
The experiment was conducted using a quantum annealing computer, the most powerful commercially available model at the time. Specifically, the D-Wave Advantage system, model Advantage system 4.1, was utilized. 
The enhanced version of this model, Advantage system 2, was not commercially available at the time of the research (March 2024). 

The computer used was characterized by the following basic technical data: a total of 5760 qubits, a Pegasus topology in a 16x16 unit cell configuration, and 40279 couplers. The number of couplers affects the ability to embed more complex problems, and therefore, with an increase in the number of couplers, the capability to sieve larger $B$-smooth numbers is expected. A more detailed specification of the device can be found in \cite{DWave}. 

The quantum computer was queried using tools from the Ocean SDK and D-Wave system library. Annealing was carried out with a sampling number of 10,000, allowing a maximum of four attempts to check the B-smoothness of a given pair. The number of samples determines the precision of the result but increases the computation time. The annealing time was set at $20 \mu s$, which is standard, well-known solution.

QUBO transforming into the QPU structure (embedding) were automatically handled using Ocean SDK tools, while the coefficients defining the specific problem were auto-scaled by the QPU solver API.

We allowed up to nine attempts to verify the smoothness of a single element, which was known in advance to be smooth. Even with nine attempts, we obtained a relatively low QPU working time, as shown below.

\subsection{Outcomes}
The most significant result achieved was the factorization of a 29-bit number: $448383577 = 20771  \times 21587$. This was accomplished with the input parameters set as follows: 
\begin{itemize}
    \item $D = 2$,
    \item $B = 224$,
    \item  $d = 4$,
    \item $W = 13$,
    \item  $M = 1$.
\end{itemize}
The polynomial $\mathcal{F}$, forming the homomorphism, was given by $$\mathcal{F} = x^4 + 2x^3 + 11x^2 + 30x + 77.$$ 
Next, after performing the sieving of smooth elements, we determined the following set: $\{(-1,1), (1,1), (1,2)\}$. These pairs represented the following elements in $\mathbb{Z}$: $-146, -144, -289$, and elements in $\mathbb{Z}[\rho]$ with the following norms: $57, 121, 1521$. By solving the system of equations, we determined the set $\mathbb{S}$, which included $\mathbb{S} = \{(1,1), (1,2)\}$. From this, we obtained the values: $X^{2} = 41616$, $\phi(\gamma) = 224202378$, and thus $gcd(448383577, 224202378-204) = 20771$.

The following results were obtained in the quantum part of the experiment. These are the largest numbers of qubits, respectively logical or physical, for all QUBO problems generated during the smoothness detection procedure for, respectively, $a+bm$ (integer part) and $N_{\mathbb{Q}(\rho)}(a+b\rho)$ (algebraic part):

\begin{itemize}
    \item logical qubits in integer part: 63,
    \item logical qubits in algebraic part: 81,
    \item physical qubits in integer part: 328,
    \item physical qubits  in algebraic part: 471,
    \item QPU working time: 980 $ \mu s$.
\end{itemize}
The "QPU working time" was summed only for the smoothness verification of the smooth elements.

 The probability of success was influenced by the choice of $N$, for which the sieved numbers $h$ were closer in size to the asymptotic values.

\section{Conclusions}

The results of the experiment address the question of the maximum practical effectiveness of solving the factorization problem using quantum annealing, assuming a hybrid combination of computations, that is, the most effective classical general method combined with a quantum QUBO solver. The decomposition of the largest problem to date, a 29-bit size, was achieved for all quantum annealing approaches, thereby improving the previous best result \cite{Zolnierczyk2023}.
 
 Although the results demonstrate new achievements, there are certain significant limitations, mainly regarding the number of steps in the quantum phase (the number of required $B$-smooth elements). The number of elements that need to be searched does not change compared to the classical method, which makes this approach unsuitable for significantly larger examples than those addressed by the best classical methods.

To better present the context of this work's result, we provide below the collected parameters of quantum methods used to factor the largest numbers. The comparison is divided into four tables: \autoref{tab1}, \autoref{tab2}, \autoref{tab3}, \autoref{tab4}, where corresponding rows refer to the same results. The general conclusions drawn from these are as follows. Quantum methods with better complexities are, so far, less practically efficient due to hardware limitations. On the other hand, hybrid methods currently demonstrate greater efficiency. We also observe that, to date, no hybrid method has been practically implemented that clearly improves the complexity compared to its classical counterpart.

\begin{table}[ht]
\centering
\caption{Factoring records: hardware used}\label{tab1}
\begin{tabular}{|c|c|c|c|}
\hline
\textbf{ID} &\textbf{Problem size (bits)} & \textbf{Hardware} & \textbf{Paper} \\
\hline
1 & 6 & gate-based quantum computer &  \cite{Amico2019}\\
\hline
2 & 10 & NMR adiabatic quantum computer & \cite{Pal2019}\\
\hline
3 & 10 & gate-based quantum computer & \cite{Selvarajan2021} \\
\hline
4 & 23 & quantum annealer &\cite{Ding2024} \\
\hline
5 & 26 & quantum annealer & \cite{Zolnierczyk2023}\\
\hline
6 & 29 & quantum annealer & this paper \\
\hline
\end{tabular}
\end{table}

\vspace{0.5cm}

\begin{table}[ht]
\centering
\caption{Factoring records: algorithms used}\label{tab2}
\begin{tabular}{|c|c|}
\hline
\textbf{ ID} & \textbf{Algorithm}  \\
\hline
1 & Shor's Algorithm  \\
\hline
2 &  Hybrid adiabatic quantum algorithm  \\
\hline
3 &  Quantum variational imaginary time evolution  \\
\hline
4 &  Quantum annealing  \\
\hline
5 &  Hybrid method with quantum annealing  \\
\hline
6 & Hybrid method with quantum annealing \\
\hline
\end{tabular}
\end{table}

\vspace{0.5cm}

\begin{table}[ht]
\centering
\caption{Factoring records: complexity}\label{tab3}
\begin{tabular}{|c|c|}
\hline
\textbf{ ID} & \textbf{Complexity} \\
\hline
1 & Polynomial \\
\hline
2 & Problem instance dependent  \\
\hline
3 & Unspecified,  \\
\hline
4 & Heuristicaly subexponential \\
\hline
5 & Heuristicaly subexponential\\
\hline
6 & Heuristicaly subexponential\\
\hline
\end{tabular}
\end{table}

\vspace{0.5cm}

\begin{table}[ht]
\centering
 l\caption{Factoring records: quantum logical resources}\label{tab4}
\begin{tabular}{|c|c|}
\hline
\textbf{ ID} & \textbf{Quantum logical resources} \\
\hline
1 & $\mathcal{O}\left(\log^{2}{N}\right)$  gate model qubits\\
\hline
2 & Unspecified \\
\hline
3 & $\mathcal{O}(\log{N})$ gate model qubits \\
\hline
4 & unspecified \\
\hline
5 &  $\mathcal{O}\left(\frac{\log^{2}{N}}{4 \left(\frac{3 \log{N}}{\log{\log{N}}}\right)^{\frac{2}{3}}}\right)$ quantum annealing qubits \\
\hline 
6 & $\mathcal{O}\left(\frac{\log^{2}{N}}{16}\right)$ quantum annealing qubits\\
\hline
\end{tabular}
\end{table}

\vspace{0.5cm}

Good effectiveness in searching for $B$-smooth numbers was observed up to a maximum size of 11-12 bits. The measured qubit cost in terms of the maximum number of logical qubits (the number of variables in the QUBO problem), equal to 81, corresponds to the asymptotic assumptions of the logical qubit cost of the applied approach: $O(\frac{\log^{2}{n}}{4})$,where $n$ is a sieved number (see \cite{Zolnierczyk2023}). The actual number of used qubits on the quantum computer, i.e., the number of physical qubits, turned out to be more than 5 times greater, due to the necessity to represent one logical qubit by many physical qubits in order to model the QUBO problem within the physical architecture of the QPU chip. 

The maximum size of the solved problem should also be related to the discrete logarithm problem. The GNFS method achieves the same size of sieved elements for the variant intended to solve the discrete logarithm problem (DLP). This means that a problem of the same size can be solved using an approach similar to that presented in this work. 

In summary, the results of the experiment should be interpreted as an answer to the question: how far are we from attacks on cryptography based on the factorization problem and DLP, which are quantum enhancements of classic general methods? Such attempts may occur sooner than the direct use of quantum methods, due to the lower resource requirements of the former approaches.




%




\bibliographystyle{splncs04}
\bibliography{BibTexBibliography.bib}

\end{document}